\documentclass[sigconf]{acmart}

\usepackage{adjustbox}
\usepackage{multirow}
\usepackage{subfigure}
\def\BibTeX{{\rm B\kern-.05em{\sc i\kern-.025em b}\kern-.08emT\kern-.1667em\lower.7ex\hbox{E}\kern-.125emX}}
    
\acmConference[IVA '19]{Intelligent Virtual Agents}{July 02--05, 2019}{Paris, France}

\begin{document}

%
\title[Perceptions of geometric features, personalities and emotions in virtual Humans]{How much do you perceive this? An analysis on perceptions of geometric features, personalities and emotions in virtual humans (Extended Version)}

%


%
\author{Victor Araujo}
\author{Rodolfo Migon Favaretto}
\author{Paulo Knob}
\author{Soraia Raupp Musse}
\affiliation{
   \institution{Graduate Program in Computer Science \\
   Pontifical Catholic University of Rio Grande do Sul}
   \city{Porto Alegre - RS}
   \state{Brazil}
}

\author{Felipe Vilanova}
\author{Angelo Brandelli Costa}
\affiliation{
   \institution{Graduate Program in Psychology \\
   Pontifical Catholic University of Rio Grande do Sul}
   \city{Porto Alegre - RS}
   \state{Brazil}
}

%
\renewcommand{\shortauthors}{Araujo, Favaretto, Knob, Musse, Vilanova and Costa.}

%
\begin{abstract}
This work aims to evaluate people's perception regarding geometric features, personalities and emotions characteristics in virtual humans. For this, we use as a basis, a dataset containing the tracking files of pedestrians captured from spontaneous videos and visualized them as identical virtual humans. The goal is to focus on their behavior and not being distracted by other features. In addition to tracking files containing their positions, the dataset also contains pedestrian emotions and personalities detected using Computer Vision and Pattern Recognition techniques. We proceed with our analysis in order to answer the question if subjects can perceive geometric features as distances/speeds as well as emotions and personalities in video sequences when pedestrians are represented by virtual humans. Regarding the participants, an amount of 73 people volunteered for the experiment. The analysis was divided in two parts: i) evaluation on perception of geometric characteristics, such as density, angular variation, distances and speeds, and ii) evaluation on personality and emotion perceptions. Results indicate that, even without explaining to the participants the concepts of each personality or emotion and how they were calculated (considering geometric characteristics), in most of the cases, participants perceived the personality and emotion expressed by the virtual agents, in accordance with the available ground truth.
\end{abstract}

%
%


%

%

%
\maketitle

\section{Introduction}
\label{sec:introduction} 

The study of human behavior is a subject of great scientific interest and probably an inexhaustible source of research~\citep{JacquesJr:2010}. Due to its importance in many applications, the automatic analysis of human behavior has been a popular research topic in the last decades~\citep{Alameda:2018}. 
In  literature, there are some work involving the visualization and analysis of cultural characteristics, such as analysis of the impact of groups on crowds through human perceptions~\cite{yang2018you}, simulation of crowds through behaviors based on personality and emotions traits~\cite{durupinar2016psychological}, visualization of interactions between virtual agents in crowd simulation and pedestrians in real video sequences~\cite{knobvisualization}, visualization of personality traits through social media~\cite{gouvisualizing}, visualization and understanding of personal emotional style~\cite{zhao2014pearl}, visualization of personal records~\cite{plaisant1996lifelines} among others. Typically, these approaches deal with Natural Language Processing (NLP) and extractions of social media data (analysis of feelings), criminal and medical records, or any other record extracted from textual data.

Recently, studies have used geometric features to analyze cultural aspects in crowds. Favaretto et al.~\cite{favaretto2016using} used group behaviors to detect cultural aspects according to Hofstede~\cite{hofstede2001culture}. In other investigations, Favaretto et al. investigated cultural aspects using controlled experiment videos (related to Fundamental Diagram~\cite{chattaraj2009comparison}) and spontaneous videos from various countries, using geometrical features~\cite{Favaretto:2016:SIB}, Big-Five personality~\cite{favaretto2017using} and OCC emotion~\cite{favaretto2018detecting} models. However, there are not many methods in the literature that investigate people's perceptions regarding geometric information~\cite{yang2018you}. In this sense, the objective of this work is to investigate how people perceive the geometric characteristics (for example, density data, distances and velocities) and non geometric characteristics (for example, cultural characteristics such as personality traits and emotions) calculated from the geometric features of pedestrians from videos of crowds. For this, we use the videos of \textit{Cultural Crowds}\footnote{(Available at: \url{http://rmfavaretto.pro.br/vhlab/})} dataset~\cite{favaretto2016using}, which contains videos of crowds from different countries, with pedestrians walking in different scenarios. Therefore, the dataset contains the tracking files with the pedestrian positions and provides also personality and emotion information of these pedestrians, which was obtained through Computer Vision and Pattern Recognition techniques. 

For the experiment, we use the track position in a simulated environment where agents were visualized as identical virtual humans. The goal is to focus on their behavior and not being distracted by other features. In our analysis the participants were asked to answer questions to identify if they can perceive geometric features as distances/speeds as well as emotions and personalities in video sequences when pedestrians are represented by virtual humans. In particularly, and very important to this work, is understand that our focus is on perception of information always related to the space and geometry, even when we talk about emotion and personality, we are interested about the pure geometric manifestations (like distance among agents, speeds and densities). The main motivation is to evaluate the area of personality and emotion detection in video sequences, i.e. we want to know if people perceive qualitatively what can be detected in video sequences.

\vspace{-0.2cm}

\section{Related Work}
\label{sec:relatedWork}
This section discusses some work related to pedestrian and crowds behavioral analysis focusing on personality traits, emotion and perception.
Knob et al.~\cite{knobvisualization} presented a work related to visualizations of interactions between pedestrians in video sequences and virtual agents in crowd simulations. Interactions are given by factors based on the OCEAN of each pedestrian and agent. The OCEAN~\cite{Digman:1990, John:1990} is the personality trait model most commonly used for this type of analysis, also referenced as Big-Five: Openness to experience (``the active seeking and appreciation of new experiences''); Conscientiousness (``degree of organization, persistence, control and motivation in goal directed behavior''); Extraversion (``quantity and intensity of energy directed outwards in the social world''); Agreeableness (``the kinds of interaction an individual prefers from compassion to tough mindedness''); Neuroticism (``how much prone to psychological distress the individual is'')~\cite{lordw07}. Durupinar et al.~\cite{durupinar2016psychological} also used OCEAN to visually represent personality traits. 
Agents' visual representation is given in several ways, for example, the animations of the agents are based on these two cultural characteristics (OCEAN and emotion). If an agent is sad, his/her animation will represent that emotion. Yang et al.~\cite{yang2018you} conducted a study on analysis perception to determine the impact of groups at various densities, using two points of view: top and first-person view. In addition to this perception, they looked at what type of camera position (top view or first-person view) could be better for the perception of density. 
The work of Ardeshir and Borji~\cite{ardeshir2018egocentric} shows experiments and graphs made between two points of view (first-person and top cam view), thus helping in the integration and use of the types of cameras used in this present work.

Regarding the detection of personalities, emotion and cultural aspects in pedestrian from crowds,~\cite{favaretto2016using} proposed a method to identify groups and characterize them to assess the aspects of cultural differences through the mapping of the Hofstede's dimensions~\cite{Hofstede:2011}. A similar idea, however using computer simulation and not focused on computer vision, is proposed by Lala et al.~\cite{Lala2012}. They use Hofstede's dimensions to create a simulated crowd from a cultural perspective. 
Gorbova and collaborators~\cite{Gorbova2017} present a system of automatic personality screening from video presentations in order to make a decision whether a person has to be invited to a job interview based on visual, audio and lexical cues. The work proposed by~\cite{favaretto2017using}, presents a model to detect personality aspects based on the Big-five personality model using individuals behaviors automatically detected in video sequences.

Several models have been developed to explain and quantify basic emotions in humans. One of the most cited is proposed by Paul Ekman~\cite{ekman1971constants} which considers the existence of 6 universal emotions based on cross-cultural facial expressions (anger, disgust, fear, happiness, sadness and surprise). 
In~\cite{favaretto2018detecting}, the authors proposed a way to detect pedestrian emotions in videos, based on OCC emotion model. To detect the emotions of each pedestrian, the authors used OCEAN as inputs, as proposed by Saifi~\cite{Saifi2016}. In our approach, we proceed with an analysis in order to verify if subjects can perceive geometric features as distances/speeds as well as emotions and personalities in video sequences when pedestrians are represented by virtual humans. Next section present how we performed the analysis.
\vspace{-0.05cm}

\section{Methodology}
\label{sec:methodology}

The main goal of this work is to analyze the perceptions of people about geometric data (speed, distance, density and angular variation), personality and emotions. The data were extracted from the Cultural Crowds dataset~\cite{favaretto2016using}. The geometric data are calculated using the pedestrian trajectories. Personality and emotion traits are also calculated based on that, through psychological hypotheses. 
Next sections detail these processes.

\subsection{Features extraction}
\label{sec:features}

Based on the tracking input file, Favaretto et al.~\cite{favaretto2017using} compute information for each pedestrian $i$ at each timestep: \textit{i)} 2D position $x_i$ (meters); \textit{ii)} speed $s_i$ (meters/frame); \textit{iii)} angular variation $\alpha_i$ (degrees) w.r.t. a reference vector $\vec{r}=(1,0)$; \textit{iv)} isolation level $\varphi_i$; \textit{v)} socialization level $\vartheta_i$; and \textit{vi)} collectivity $\phi_i$. To compute the collectivity affected in individual $i$ from all $n$ individuals, they computed $\phi_i = \sum_{j=0}^{n-1} \gamma e^{(-\beta \varpi(i,j)^{2})}$, and the collectivity between two individuals was calculated as a decay function of $\varpi(i,j) = s(s_i,s_j).w_1+o(\alpha_i,\alpha_j).w_2$, considering $s$ and $o$ respectively the speed and orientation differences between two people $i$ and $j$, and $w_1$ and $w_2$ are constants that should regulate the offset in meters and radians.

To compute the socialization level $\vartheta$, Favaretto et al.~\cite{Favaretto:2016:SIB} use an artificial neural network (ANN) with a Scaled Conjugate Gradient (SCG) algorithm in the training process to calculate the socialization $\vartheta_i$ level for each individual $i$. The ANN has 3 inputs (collectivity $\phi_i$ of person $i$, mean Euclidean distance from a person $i$ to others $\bar{d_{i,j}}$ and the number of people in the Social Space\footnote{Social space is related to $3.6$ meters~\cite{hall98}.} according to Hall's proxemics~\cite{hall98} around the person $n_i$). The isolation level corresponds to its inverse, $\varphi_i = 1 - \vartheta_i$. For more details about how this features are obtained, please refer to~\cite{favaretto2017using, Favaretto:2016:SIB}. For each individual $i$ in a video, we computed the average for all frames and generate a vector $\vec{V_i}$ of extracted data where $\vec{V_i} = \left [x_i, s_i, \alpha_i, \varphi_i, \vartheta_i, \phi_i \right ]$. In the next section we describe how these features are mapped into personality  and emotion traits.

\subsection{Personality and emotion detection}
\label{sec:pers_emo}

To detect the five dimensions of OCEAN for each pedestrian,~\cite{favaretto2017using} used the NEO PI-R~\cite{Costa:1992} that is the standard questionnaire measure of the Five Factor Model. They firstly selected NEO PI-R items related to individual-level crowd characteristics and the corresponding OCEAN-factor. For example: "Like being part of crowd at sporting event" corresponding to the factor "Extroversion". As described in details in~\cite{favaretto2017using}, they proposed a series of empirically defined equations to map pedestrian features to OCEAN dimensions. Firstly, they selected 25 from the 240 items from NEO PI-R inventory that had a direct relationship with crowd behavior. In order to answer the items with data coming from real video sequences, they proposed equations that could represent each one of the 25 items with features extracted from videos. For example, in order to represent the item ``1 - Have clear goals, work to them in orderly way'', Favaretto and his colleagues consider that the individual $i$ should have a high velocity $s$ and low angular variation $\alpha$ to have answer in concordance with this item. So the equation for this item was $Q_1 = s_i+\frac{1}{\alpha_i}$. In this way, they empirically proposed equations for all the 25 items, as presented in~\cite{favaretto2017using}.

In the work presented by~\cite{favaretto2018detecting}, the authors proposed a way to map OCEAN dimensions of each pedestrian in OCC Emotion model, regarding four emotions: Anger, Fear, Happiness and Sadness. This mapping is described in Table~\ref{tab:emotionMapping}. In Table~\ref{tab:emotionMapping}, the plus/minus signals along each factor represent the positive/negative value of each one. For example concerning Openness, O+ stands for positive values (i.e. O $\geq$ 0.5) and O- stands for negative values (i.e. O $<$ 0.5)). A positive value for a given factor (i.e. 1) means the stronger the OCEAN trait is, the stronger is the emotion too. A negative value (i.e. -1) does the opposite, therefore, the stronger the factor's value, the weaker is a given emotion. A zero value means that a given emotion is not affected at all by the given factor.

\begin{table}[htb]
   \footnotesize
   \centering
   \caption{Emotion mapping from OCEAN to OCC~\cite{favaretto2018detecting}.}
   \begin{adjustbox}{max width=\textwidth}
     \begin{tabular}{c|cccc}
     \hline\noalign{\smallskip}
     \textbf{Factor} & \textbf{Fear} & \textbf{Happiness} & \textbf{Sadness} & \textbf{Anger} \\
     \noalign{\smallskip}
        \hline
     \noalign{\smallskip}
     O+ & 0 & 0 & 0 & -1 \\
     O- & 0 & 0 & 0 & 1 \\
     C+ & -1 & 0 & 0 & 0 \\ 
     C- & 1 & 0 & 0 & 0 \\ 
     E+ & -1 & 1 & -1 & -1 \\ 
     E- & 1 & 0 & 0 & 0 \\ 
     A+ & 0 & 0 & 0 & -1 \\ 
     A- & 0 & 0 & 0 & 1 \\ 
     N+ & 1 & -1 & 1 & 1 \\ 
     N- & -1 & 1 & -1 & -1 \\ 
     \noalign{\smallskip}
        \hline
     \end{tabular}
   \end{adjustbox}
   \label{tab:emotionMapping}
\end{table}

\vspace{-0.2cm}

\subsection{Features visualization}
\label{sec:viwer}

\begin{figure}[!htb]
    \centering
    \includegraphics[width=0.7\linewidth]{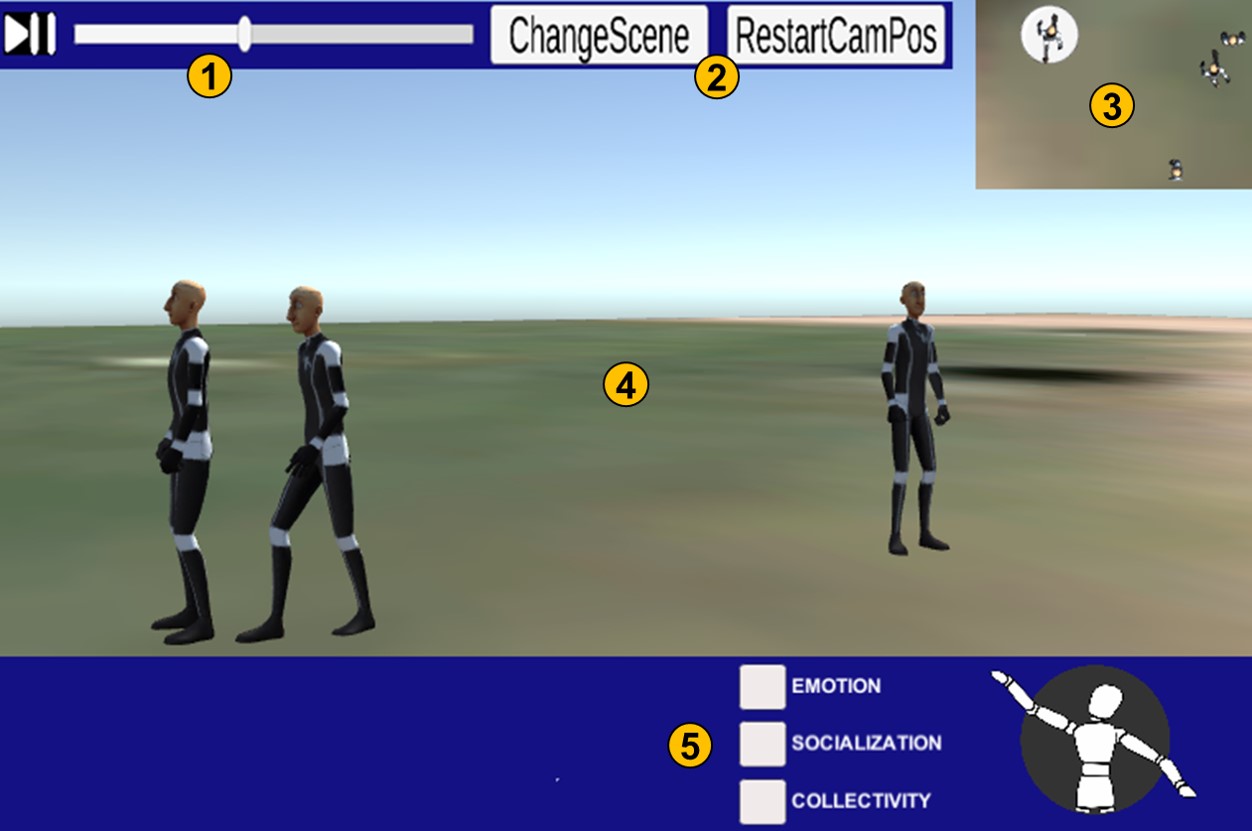}
    \caption{Main window of the viewer.}
    \label{fig:viewer_cam_1p}
\end{figure}

The viewer was developed using the Unity3D\footnote{Unity3D is available at \url{https://unity3d.com/}} engine, with $C\#$ programming language. The viewer allows the users to rewind, accelerate and stop the simulated video through a time controller, so that the user can observe something that he/she finds interesting several times, at any time. Figure~\ref{fig:viewer_cam_1p} shows the main window of the viewer. As identified in Figure~\ref{fig:viewer_cam_1p}, the viewer is divided in five parts, as follows: 1) time controller, where is possible to start, stop and continue simulation playback; 2) buttons \textit{ChangeScene} and \textit{RestartCamPos} to, respectively, load the data file of another video and restart the camera position for viewing in first person; 3) a window that shows the top view of the environment; 4) the first-person view of a previously selected agent (this agent is highlighted in area 3) and  5) that contains features panel, where the users can activate the visualization of the data related to the emotion, socialization and collectivity of agents.

\begin{figure*}[!h]
  \centering
  \subfigure[fig:camerasFig][Top view]{\includegraphics[width=0.335\textwidth]{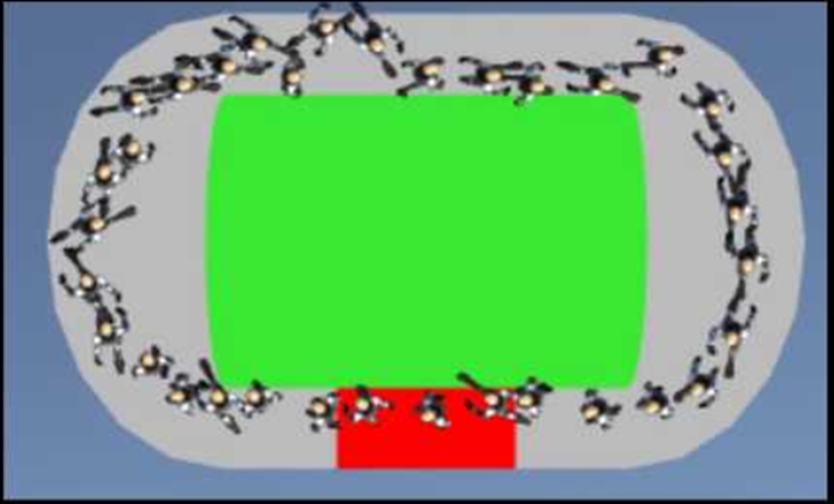}}
  \subfigure[fig:camerasFig][Oblique view]{\includegraphics[width=0.335\textwidth]{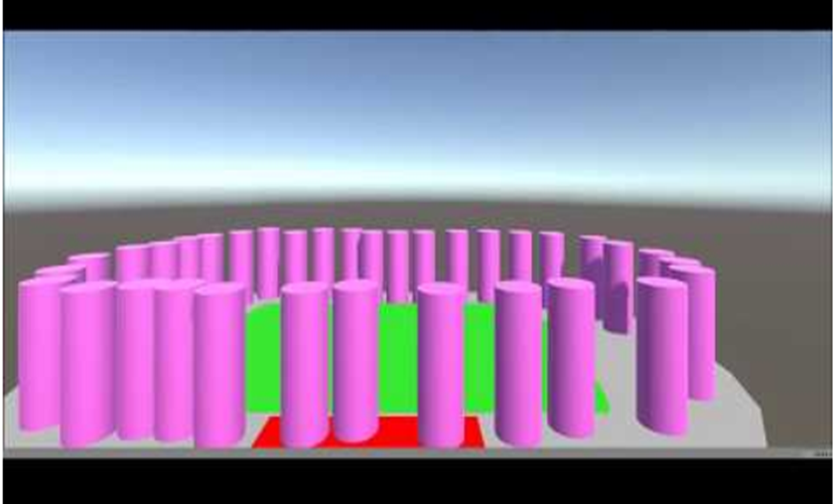}}
  \subfigure[fig:camerasFig][First-person view]{\includegraphics[width=0.32\textwidth]{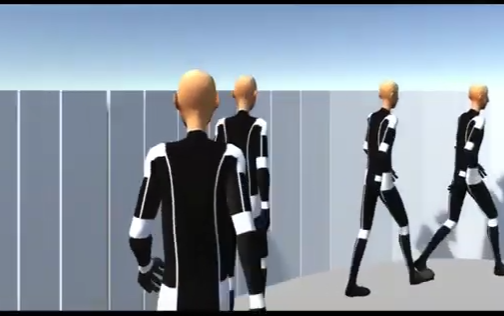}}
    \caption{Types of visualization: (a) top view, (b) oblique and (c) first-person view.}
    \label{fig:camerasFig}
\end{figure*}
This viewer has three modes of visualization: (i) first-person visualization, 
(ii) top view, and 
(iii) an oblique view. 
Figure~\ref{fig:camerasFig} shows an example of each type of camera point of view in a video 
available in the Cultural Crowds dataset. In addition to these different points of view, it is possible to observe all the pedestrians present in each frame $f$. Pedestrians can be represented by an humanoid or cylinder type avatar. Each pedestrian $i$ present in frame $f$ has a position ($X_i,Y_i$) (already converted from image coordinates to world coordinates). In addition to the positions, it is also possible to know if the pedestrian is walking, running or stopped in frame $f$ through the current speed $s_i$. If in this frame the current speed is greater than or equal to $\frac{0.08m}{f}$ which is equivalent to $\frac{2m}{s}$, considering $\frac{24f}{s}$, then the avatar is running. It was defined based on the Preferred Transition Speed PTS~\cite{alexander1992model}. The values of the transitions can be seen in Equation~\ref{eq:animationTransition}, considering the current speed of the agent $s_i$.
\begin{equation}
Animation = \left\{
\begin{array}{ll}
    \textbf{Idle}, & \mbox{when $s_i == 0$}; \\
    \textbf{Walk}, & \mbox{when $0  < s_i <  \frac{0.08m}{f}$}; \\
    \textbf{Run}, & \mbox{when $s_i \geq \frac{0.08m}{f}$}.
\end{array} \right . 
\label{eq:animationTransition}
\end{equation}
Also, for the humanoid avatar type, each speed transition is accompanied by an animation transition, for example, if the current speed $s_i == 0$, then it does not change the animation (remaining stationary), but if its speed is $0  < s_i < \frac{0.08m}{f}$, then the animation changes for walking as well as if $s_i \geq \frac{0.08m}{f}$, the animation of the avatar changes to running. Next section presents some obtained results.

\section{Results}
\label{sec:results}

This section aims to present the results of people's perceptions about geometric data information (density, speed, distance between pedestrians and angular variation), personalities and emotions. We used the simulation environment to generate some short sequences of pedestrian videos together with a questionnaire where the sequences of videos are presented. In the sequence, participants' responses were analyzed. This section was organized into three parts: Section~\ref{sec:visualizatorResults} presents some information about the videos from the dataset that were used in the experiment, Section~\ref{sec:questionnaireResults} discuss the results of the perceptions about the geometrical characteristics of pedestrians and Section~\ref{sec:perceptionsResults} presents the results of the perceptions about personalities and emotions.

\subsection{Video characteristics}
\label{sec:visualizatorResults}

We generate video sequences with data extracted from the \textit{Cultural Crowds} dataset. Table~\ref{tbl:dataVideos} shows the relations of all videos from the dataset that were used in the experiment, with information about the country where the video was recorded, the number of pedestrians and the density level (low, medium, or high). The data of each chosen video was input to a simulated environment containing virtual agents, represented by cylinder or humanoid type avatars, that can be seen, respectively, in Figure~\ref{fig:camerasFig}(b) and (c). We also used the three point of view cameras (top view illustrated in Figure~\ref{fig:camerasFig}(a)). Regarding the participants, an amount of 73 people volunteered for the experiment: 45 males (61.6\%) and 28 (38.4\%) females and 47.9\% have some undergraduate degree. In the next section we discuss the results obtained in the geometrical features perception analysis.
\begin{table}[!htb]
    \footnotesize
    \centering
    \caption{Videos of the Cultural Crowds~\cite{favaretto2016using} dataset.}
    \label{tbl:dataVideos}
    \begin{tabular}{c|c|c|c}
        \hline
        \noalign{\smallskip}
        Video & Country & N. Pedestrian & Density \\ 
        \noalign{\smallskip}
        \hline
        \noalign{\smallskip}
        \hline
        \noalign{\smallskip}
        AE-01 & Unit. Arab Emirates & 12 & Low \\
        AT-03 & Austria & 10 & Low \\ 
        BR-01 & Brazil  & 16 & Low \\ 
        BR-15 & Brazil & 15 & Low \\ 
        BR-25 & Brazil & 25 & Medium \\ 
        BR-34 & Brazil & 34 & High \\
        \noalign{\smallskip}
        \hline
    \end{tabular}
\end{table}


\subsection{Geometric features perception}
\label{sec:questionnaireResults}
\begin{figure}[!htb]
    \centering
    \includegraphics[width=0.9\linewidth]{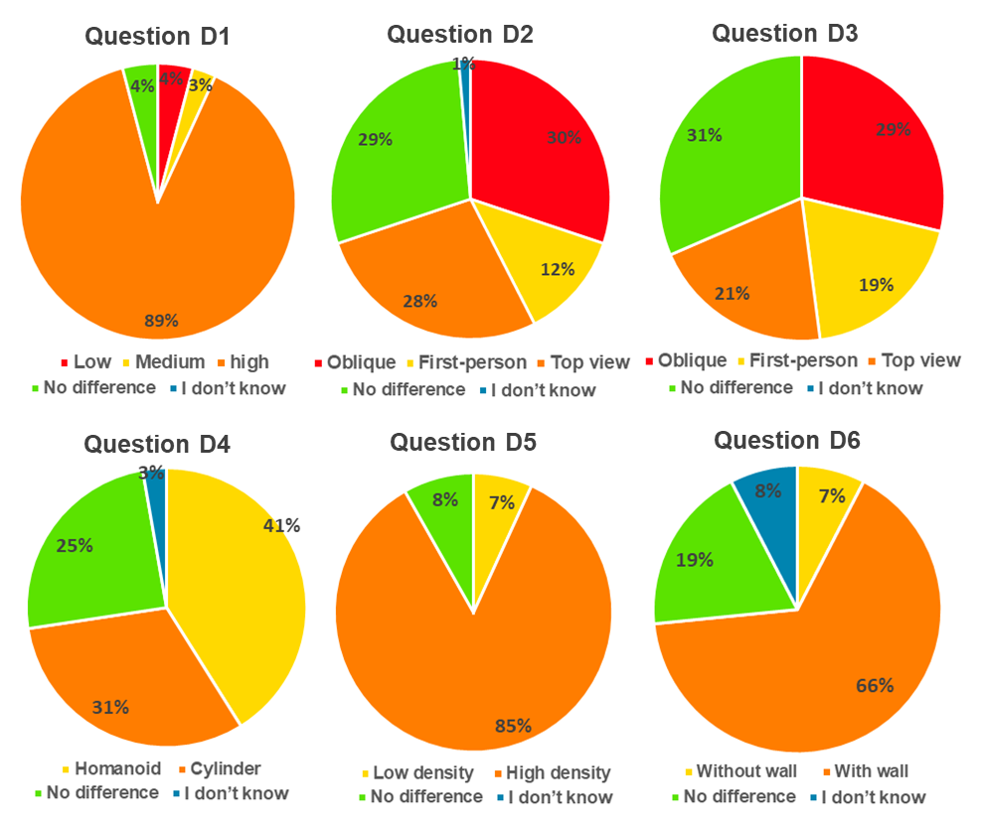}
    \caption{Perception concerning density: questions D1 to D6.}
    \label{fig:perception_density}
\end{figure}

In this section, we present an analysis of subjects perception regarding density, velocity,  direction variation of pedestrians and distance among them using three camera's points of view (first-person, oblique and top-view) and two types of avatars (cylinder and humanoid). 
The first part of applied questionnaire contains six questions but in all of them we asked for the same aspect: "In which video do you perceive the higher density?". Before each question, two or three short videos described in Table~\ref{tbl:dataVideos} were presented. Figure~\ref{fig:perception_density} shows the questions and percentage of answers.

\begin{figure}[!htb]
    \centering
    \includegraphics[width=0.9\linewidth]{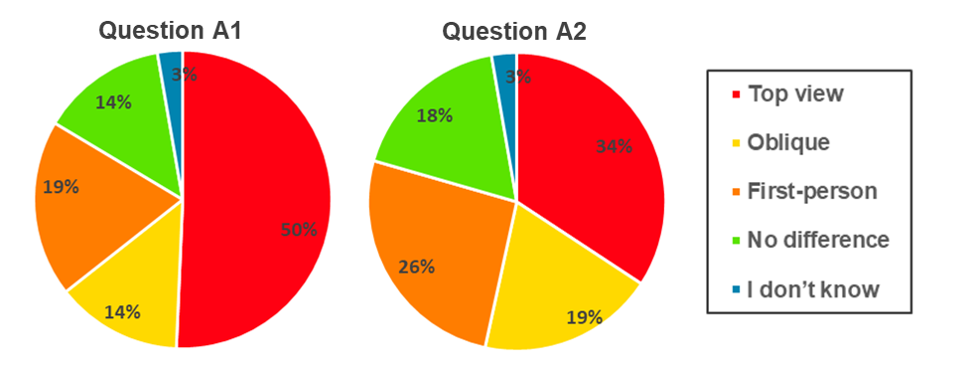}
    \caption{Perception analysis concerning angular variation: questions A1 and A2.}
    \label{fig:perception_angvar}
\end{figure}

The first question (D1) aimed to evaluate if the participants can perceive the density variation once we did not include any explanation about that. Therefore, scenes of videos with low, medium and high density of people in crowds were presented where we want to to check if the subjects could correctly select the high density one.
89\% of participants responded according to ground truth, i.e. they could correctly classify  the high density video. The other 11\% answered "I do not know", "I did not notice density difference" and low and medium density options.
In D2 and D3 we presented videos with same density but displayed with the different points of view, however in D2 we used humanoids and in D3 we used cylinders. We asked to the subjects to select the video where the higher density was observed. Our goal was to check if the subjects could perceive the same density or if density perception changes due to the camera point of view or the way the agents are displayed.
In question D2, 
70\% chose one of the videos, while 29\% of the participants marked the option "I did not notice density difference", so for this small group it seems that the camera does not change the perception. Details are presented in Figure~\ref{fig:perception_density}, and results
indicate that the camera point of view can disturb the density perception. Regarding the point of view, oblique cameras present the higher percentage of answers. In question D3, 69\% chosen one of the videos, while 31\% of the participants marked the option "I did not notice density difference", indicating that the visualization with cylinders or humanoids also change the final result. In question D4, we showed two videos with same density and same point of view, however changing the type of avatar. 25\% of people selected the option "I did not notice density difference", while 72\% chosen one of the avatar types, being 41\% of the participants have chosen humanoids.
In questions D5 and D6, we included, in same videos analyzed before, walls that surround the agents (see Figure~\ref{fig:camerasFig}(c)). The goal is to check if it changes the density perception using first-person camera. In this case 66\% of subjects answered that one of the videos presented higher density in comparison to a same density video without walls.
Regarding speed perception, the questionnaire also contains six questions, all of them are related to low-density videos described in Table~\ref{tbl:dataVideos}. The goal of these questions is to evaluate the speed levels running and walking, as presented in Equation~\ref{eq:animationTransition}, through the top and oblique cameras, in addition to the two types of avatars: cylinder and humanoid.
\begin{figure}[!htb]
    \centering
    \includegraphics[width=0.9\linewidth]{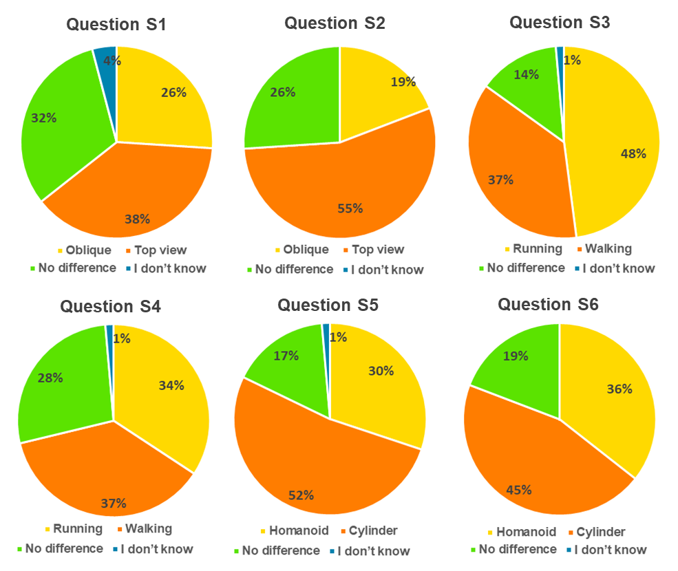}
    \caption{Perception concerning speed: questions S1 to S6.}
    \label{fig:perception_speed}
\end{figure}%
In such videos there was no analysis of perceptions using the camera in the first person, since we observe that such videos did not allow a good vision of the scene. 
As in density analysis, we asked the same question "In which video did you observe the higher velocity" and showed variations of parameters we want to measure. Question S1 presented two videos with velocity=running and cameras=oblique and top. As shown in Figure~\ref{fig:perception_speed}, 32\% of subjects do not perceive any difference in velocity while 64\% chosen one of the videos. Same process for question S2 but using velocity=walking and 26\% does not perceive difference while 74\% chosen one of the videos. Questions S3 and S4 presented same velocity respectively in oblique and top of view camera. For S3, 14\% of subjects do not perceive velocity changes while 85\% selected only one of the cameras. In S4, 28\% of them do not perceive velocity changes while 71\% selected only one of the cameras. Finally questions S5 and S6 presented two videos containing the two different avatars with oblique and top camera respectively for velocity=walking. Results were very similar having 17\% and 19\% respectively of people who do not perceive difference against 82\% and 81\% of people that chose one of the videos. So, our results indicate that the camera point of view and type of avatar impacts in the velocity perception.
Regarding the perception of angular variation, the questionnaire contains two questions with comparisons between the three types of cameras and two types of avatars. All angular variation questions used scenes from BR-34 (high density) video shown in Table~\ref{tbl:dataVideos}. 
Again, we asked the same question "In which video do you observe more angular variation performed by the agents?" and videos variate the measured parameters. Question A1 presented three videos with humanoids viewed with 3 different camera positions. As shown in Figure~\ref{fig:perception_angvar}, only 14\% of subjects do not perceive difference in the angular variation while 83\% chosen one of the videos and the top view camera was more selected. Similar process for question A2 where avatars were cylinders. 18\% of subjects did not perceive difference while 79\% selected one of the videos. Most part of people who selected one video chosen the one with humanoids.

\begin{figure}[!htb]
    \centering
    \includegraphics[width=0.9\linewidth]{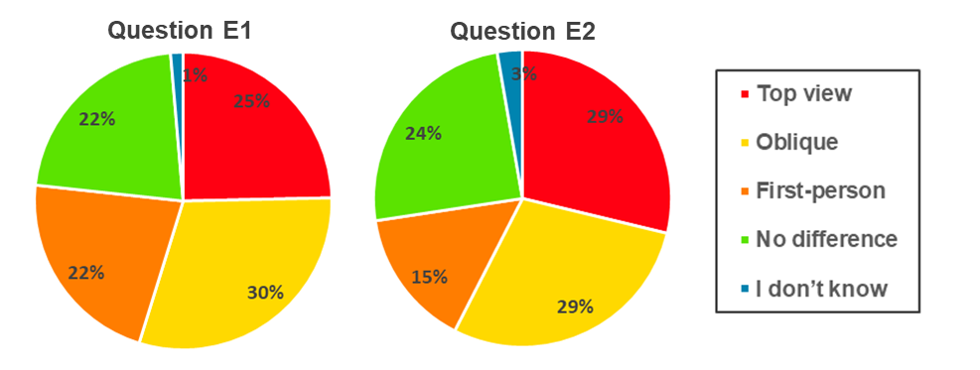}
    \caption{Perception concerning distance: questions E1 and E2.}
    \label{fig:perception_distance}
\end{figure}

Regarding the perception of distance between the avatars, the questionnaire contains two questions, all with videos containing high density. The videos used in these questions were the same as the questions about the perception of angular variation, i.e the types of cameras and the two types of avatars, and the question is: "In which video do you observe the largest distance among agents?". 
Indeed, results were very similar in both question E1 and question E2. As shown in Figure~\ref{fig:perception_distance}, in E1 we displayed humanoids with the three cameras and 22\% of subjects do not perceive differences, while in E2 we displayed cylinders and 24\% also do not perceive changes. On the other hand, 77\% and 73\% of subjects, respectively, selected one of the videos in a approximately uniformly distributed way.

So, in this section we analyzed the subjects perception related to density, speed, angular variation and distances among agents displayed using two types of avatars and in three different cameras point of view. Results indicate that changing the way we displayed avatars and cameras position the subjects perception also changes. In particular, top of view and oblique cameras seem to provide better information to detect the parameters while humanoids were preferred to indicate the higher values of all evaluated parameters.


\subsection{Personality and emotion perceptions}
\label{sec:perceptionsResults}

In this section we present the part of this study focused on perception of personality and emotion traits in crowd videos. As explained before, we used the simulation environment to generate some short sequences of pedestrian views in low density crowds (due to the data present in the dataset). In each video sequence we highlighted two individuals with different colors (red and yellow) and we asked to the subjects about them. Table~\ref{tab:survey_perception} shows the questions with the possible answers, where the correct answer of each question is highlighted in bold. We use as ground truth the results obtained by the approach proposed by Favaretto et al.~\cite{favaretto2018detecting}. 

\begin{table}[!htb]
    \centering
    \caption{Questions and possible answers. The correct answer in highlighted in bold, according to Favaretto et al.~\cite{favaretto2018detecting}.}
    \begin{adjustbox}{max width=0.85\linewidth}
        \begin{tabular}{p{8.5cm}|l}
            \hline\noalign{\smallskip}
            Question & Possible answers \\
            \noalign{\smallskip}
            \hline
            \noalign{\smallskip}
            \hline
            \noalign{\smallskip}
            \multirow{5}{*}{\shortstack[l]{\textbf{Q1}: In your opinion, which of the two pedestrians highlighted in \\ the video has a neurotic personality, yellow or red?}}
            & \textit{a}) Yellow pedestrian; \\
            & \textit{b}) \textbf{Red pedestrian}; \\
            & \textit{c}) Both pedestrians; \\
            & \textit{d}) Neither of them; \\
            & \textit{e}) I don't know. \\
            \noalign{\smallskip}
            \hline
            \noalign{\smallskip}
            \multirow{5}{*}{\shortstack[l]{\textbf{Q2}: In your opinion, which of the two pedestrians highlighted in \\ the video is angry, yellow or red?}}
            & \textit{a}) Yellow pedestrian; \\
            & \textit{b}) \textbf{Red pedestrian}; \\
            & \textit{c}) Both pedestrians; \\
            & \textit{d}) Neither of them; \\
            & \textit{e}) I don't know. \\
            \noalign{\smallskip}
            \hline
            \noalign{\smallskip}
            \multirow{5}{*}{\shortstack[l]{\textbf{Q3}: In your opinion, which of the two pedestrians highlighted in \\ the video is more openness to experiences, yellow or red?}}
            & \textit{a}) \textbf{Yellow pedestrian}; \\
            & \textit{b}) Red pedestrian; \\
            & \textit{c}) Both pedestrians; \\
            & \textit{d}) Neither of them; \\
            & \textit{e}) I don't know. \\
            \noalign{\smallskip}
            \hline
            \noalign{\smallskip}
            \multirow{5}{*}{\shortstack[l]{\textbf{Q4}: In your opinion, which of the two pedestrians highlighted in \\ the video is afraid, yellow or red?}}
            & \textit{a}) Yellow pedestrian; \\
            & \textit{b}) \textbf{Red pedestrian}; \\
            & \textit{c}) Both pedestrians; \\
            & \textit{d}) Neither of them; \\
            & \textit{e}) I don't know. \\
            \noalign{\smallskip}
            \hline
            \noalign{\smallskip}
            \multirow{5}{*}{\shortstack[l]{\textbf{Q5}: In your opinion, which of the two pedestrians highlighted in \\ the video is happier, yellow or red?}}
            & \textit{a}) \textbf{Yellow pedestrian}; \\
            & \textit{b}) Red pedestrian; \\
            & \textit{c}) Both pedestrians; \\
            & \textit{d}) Neither of them; \\
            & \textit{e}) I don't know. \\
            \noalign{\smallskip}
            \hline
            \noalign{\smallskip}
            \multirow{5}{*}{\shortstack[l]{\textbf{Q6}: In your opinion, which of the two pedestrians highlighted in \\ the video is more extroverted, yellow or red?}}
            & \textit{a}) \textbf{Yellow pedestrian}; \\
            & \textit{b}) Red pedestrian; \\
            & \textit{c}) Both pedestrians; \\
            & \textit{d}) Neither of them; \\
            & \textit{e}) I don't know. \\
            \noalign{\smallskip}
            \hline
            \noalign{\smallskip}
            \multirow{5}{*}{\shortstack[l]{\textbf{Q7}: In your opinion, which of the two pedestrians highlighted in \\ the video seems to be more sociable, yellow or red?}}
            & \textit{a}) \textbf{Yellow pedestrian}; \\
            & \textit{b}) Red pedestrian; \\
            & \textit{c}) Both pedestrians; \\
            & \textit{d}) Neither of them; \\
            & \textit{e}) I don't know. \\
            \noalign{\smallskip}
            \hline
         \end{tabular}
     \end{adjustbox}
   \label{tab:survey_perception}
\end{table}

\begin{figure}[!htb]
    \centering
    \subfigure[Initial frame]{\includegraphics[width=0.49\linewidth]{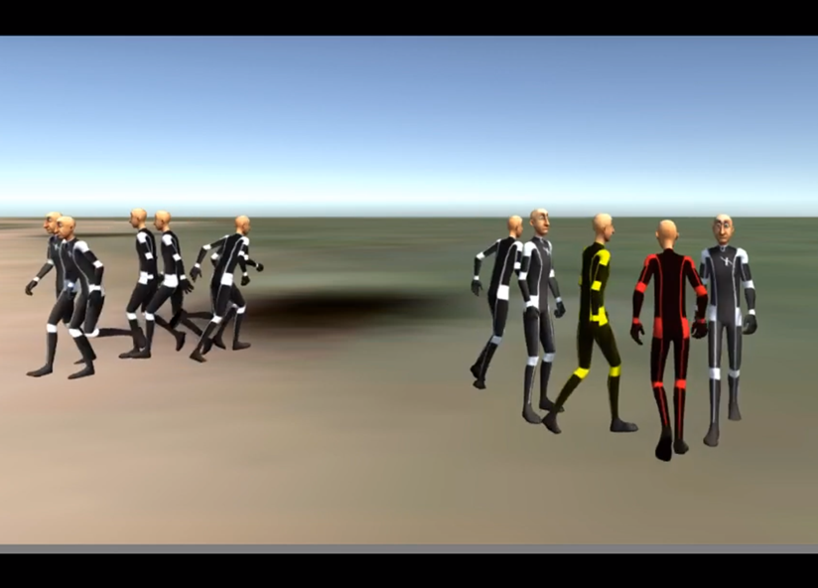}}\hfill
    \subfigure[Final frame]{\includegraphics[width=0.49\linewidth]{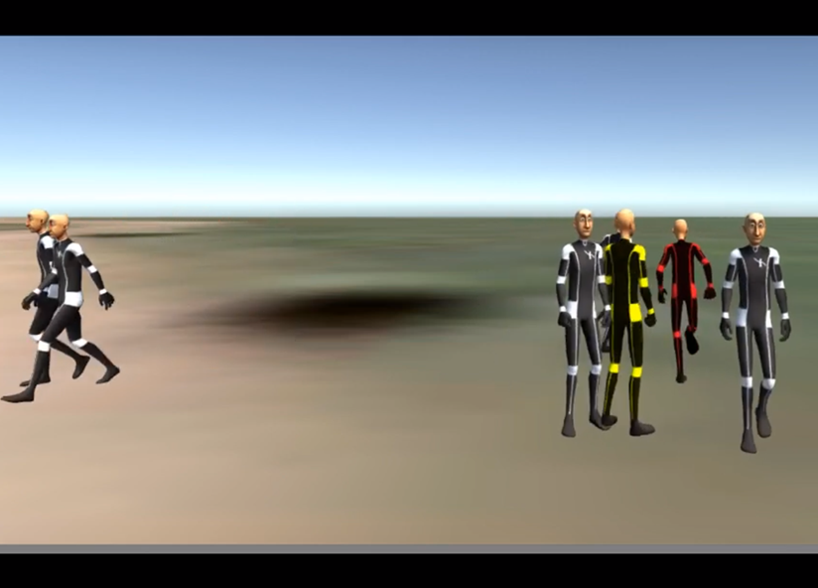}}
    \caption{Initial (a) and final (b) frames from video $P01$.}
    \label{fig:perception_P01}
\end{figure}

\begin{figure}[!htb]
    \centering
    \includegraphics[width=0.95\linewidth]{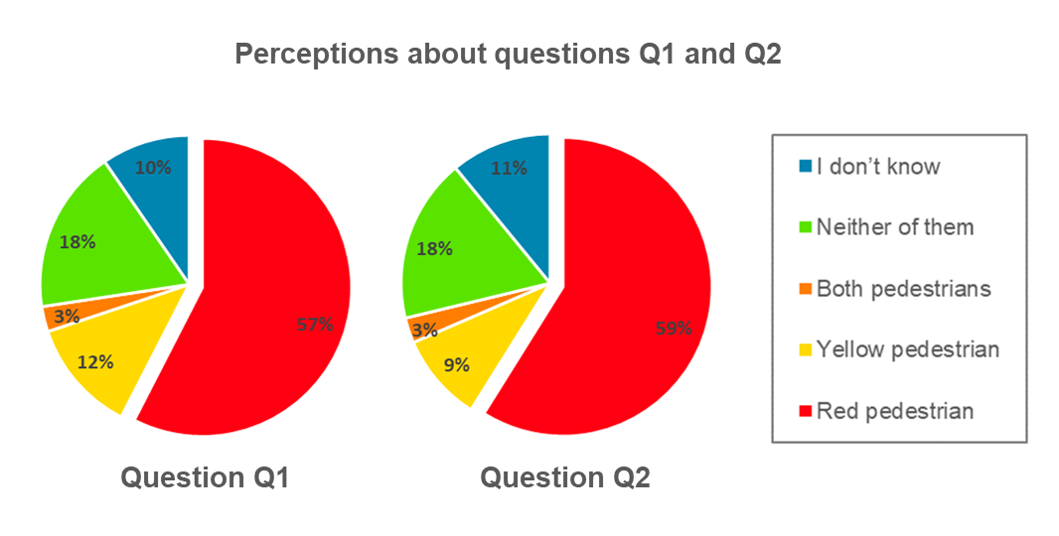}
    \caption{Perception analysis concerning Q1 and Q2.}
    \label{fig:results_P01}
\end{figure}

Figure~\ref{fig:perception_P01} shows the initial and final frames from the video $P01$, where it is possible to see a group of pedestrians in the right part of the video. Pedestrian highlighted in yellow is part of this group and the pedestrian highlighted in red walk trough the group with a higher speed. In the questions $Q1$ and $Q2$ (related to the video $P01$) we asked about which pedestrian (yellow or red) was, respectively, neurotic and angry. Figure~\ref{fig:results_P01} shows the answers given by the participants. It was interesting to see that a little bit more than half of participants (57\% in $Q1$ and 59\% in Q2) answered according to the ground truth. The pedestrian highlighted in red was the most neurotic and angry, according to Favaretto's approach. Only a few participants answered that the pedestrian highlighted in yellow was neurotic and scared (12\% in question $Q1$ and 9\% in question $Q2$) and 18\% answered that ``neither of them'' was neurotic. As proposed by~\cite{favaretto2018detecting}, geometrically, a neurotic person remains isolated and few collective. So, subjects who do think that no agent was neurotic was certainly thinking about the psychological point of view, while we are analyzing based on space relationship. In video $P01$, the pedestrian highlighted in red has these characteristics. The pedestrian highlighted in red is angry: isolated, low angular variation, low speed, low socialization and low collectivity.

\begin{figure}[!htb]
    \centering
    \subfigure[Initial frame]{\includegraphics[width=0.49\linewidth]{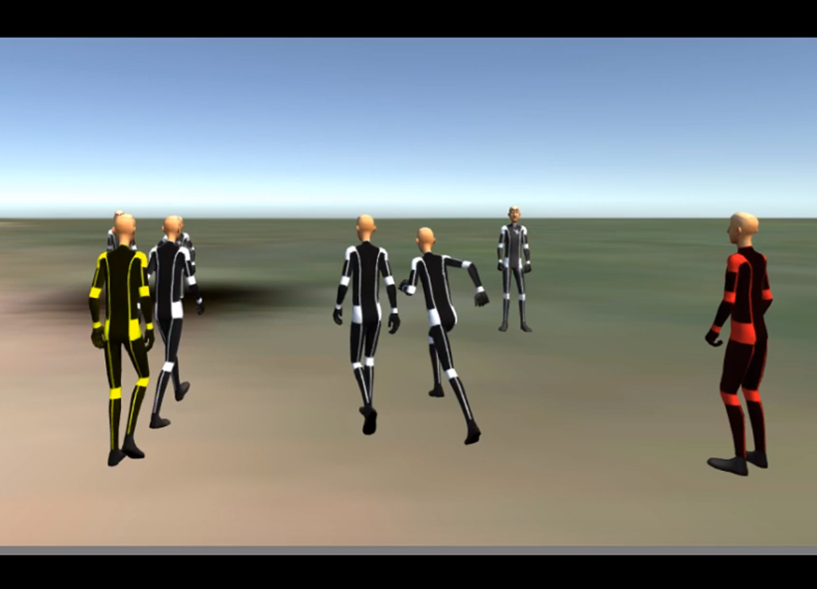}}\hfill
    \subfigure[Final frame]{\includegraphics[width=0.49\linewidth]{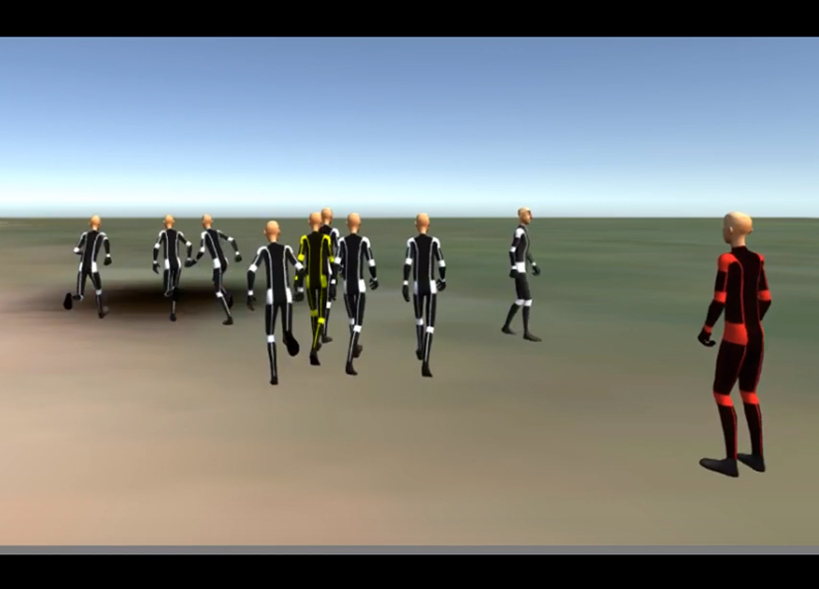}}
    \caption{Initial (a) and final (b) frames from video $P02$.}
    \label{fig:perception_P02}
\end{figure}

\begin{figure}[!htb]
    \centering
    \includegraphics[width=0.95\linewidth]{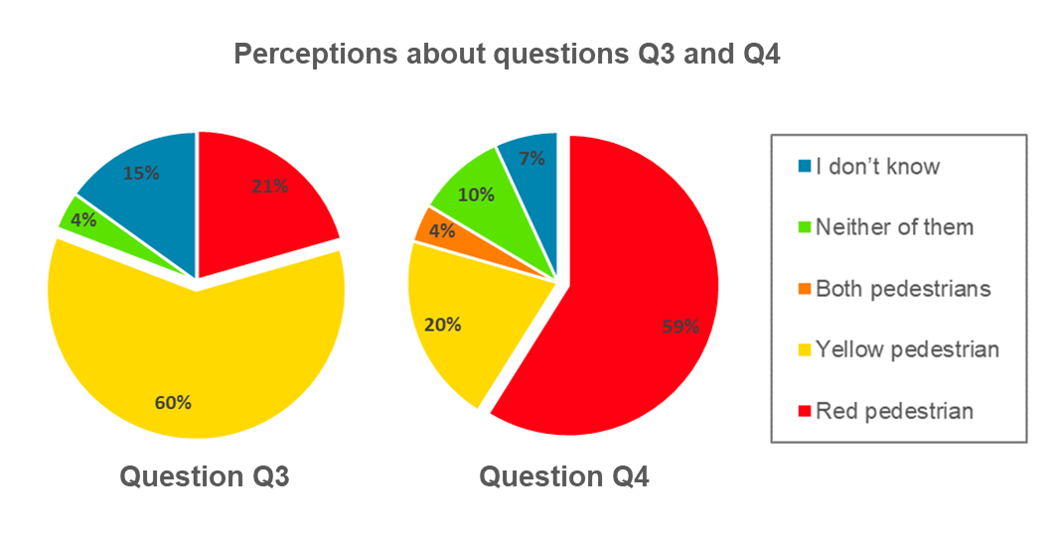}
    \caption{Perception analysis concerning Q3 and Q4.}
    \label{fig:results_P02}
\end{figure}

Following the analysis, video $P02$ (illustrated in Figure~\ref{fig:perception_P02}) has a pedestrian highlighted in yellow interacting with a group of individuals and a pedestrian highlighted in red who is alone and not interacting with anyone. Questions $Q3$ and $Q4$, who were related to this video, asked participants about which highlighted pedestrian was, respectively, openness to experiences and afraid.  Figure~\ref{fig:results_P02} shows the answers for that questions. The results plotted in Figure~\ref{fig:results_P02} shows that most of the participants perceived the same personality (in case of question $Q3$) an the same emotion (question $Q4$) when compared to ground truth, i.e.  60\% of the participants correctly chose the pedestrian in yellow as the most opened to experiences in question $Q3$ and 59\% correctly chose the pedestrian in red as having fear. In the model of~\cite{favaretto2018detecting}, a pedestrian opened to new experiences is related to a high value for the angular variation feature. Geometrically speaking, according to what has been proposed in our model, a person who allows himself/herself to change  objectives (direction) while walking is more subject to new experiences. Fear, in turn, is linked to the fact that the person is isolated from others and  walks at lower speeds.

\begin{figure}[!htb]
    \centering
    \subfigure[Initial frame]{\includegraphics[width=0.49\linewidth]{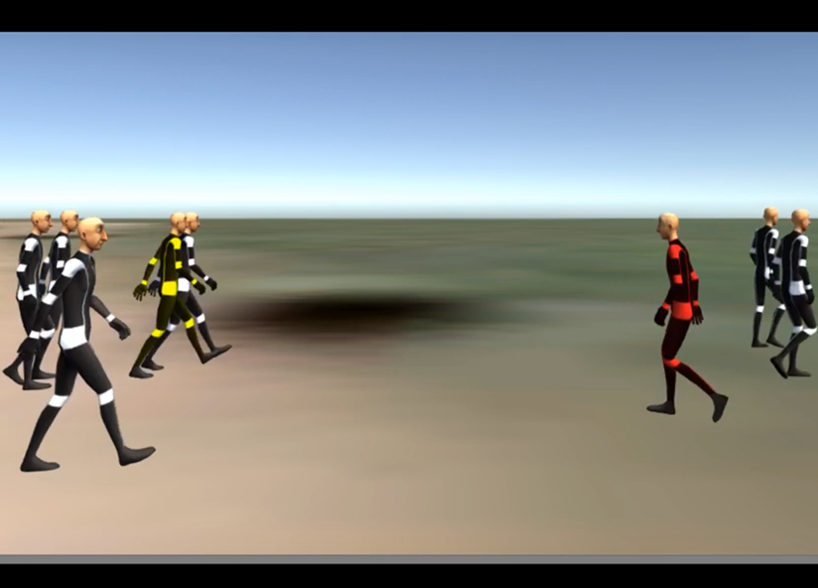}}\hfill
    \subfigure[Final frame]{\includegraphics[width=0.49\linewidth]{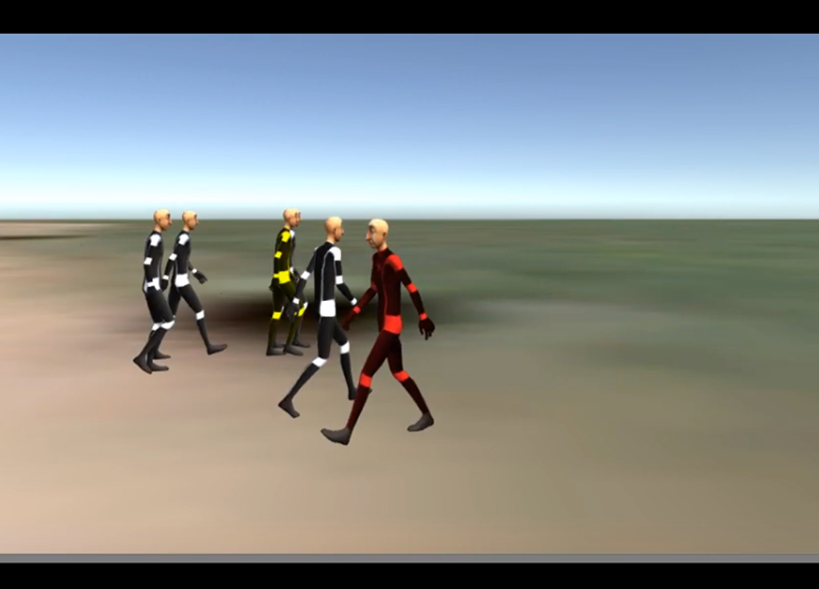}}
    \caption{Initial (a) and final (b) frames from video $P03$.}
    \label{fig:perception_P03}
\end{figure}

\begin{figure}[!htb]
    \centering
    \includegraphics[width=0.95\linewidth]{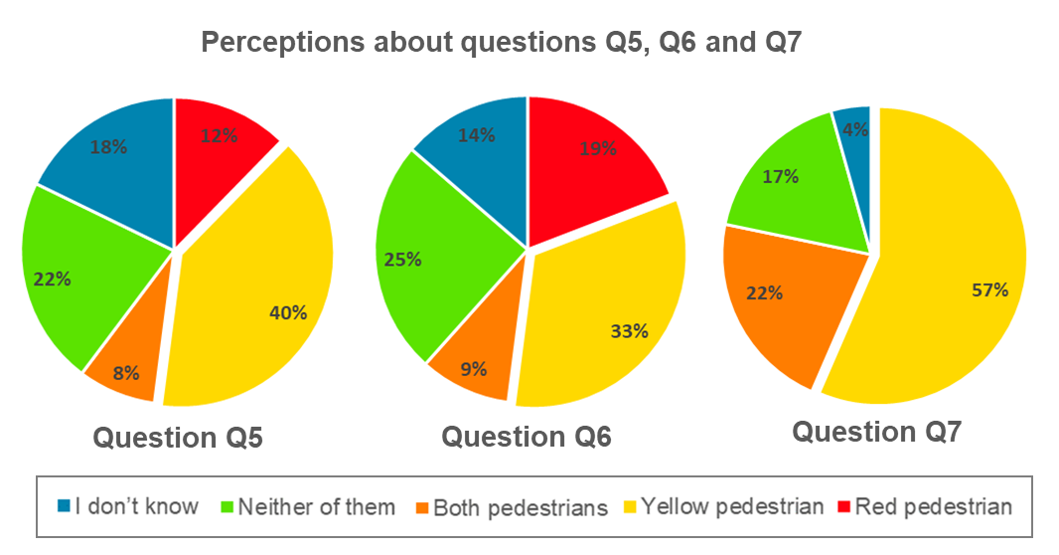}
    \caption{Perception analysis concerning Q5, Q6 and Q7.}
    \label{fig:results_P03}
\end{figure}

Finally, related to the video $P03$
we propose questions Q5, Q6 and Q7, asking, respectively, about happiness, extraversion and sociability. The video $P03$ (illustrated in Figure~\ref{fig:perception_P03}) contains a pedestrian highlighted in yellow walking with a group of people and a pedestrian highlighted in red walking alone, in the opposite direction of all other pedestrians. Regarding question $Q5$ (plotted on the left side of Figure~\ref{fig:results_P03}), 40\% of participants answered according to the ground truth. Geometrically, a happy person is not isolated and can present high levels of collectivity and socialization. Pedestrian highlighted in yellow presented that characteristics and was correctly identified by the participants in the survey.

Questions $Q6$ and $Q7$ analyze, respectively, extraversion and sociability. In question $Q6$, although most of the participants (33\% of them) correctly answer that the pedestrian highlighted in yellow is the most extrovert, it seems that the participants were not very sure about perceiving this characteristic. 25\% of them answered that none of the pedestrians were extroverted, 19\% replied that the most extroverted pedestrian was the one highlighted in red, 14\% did not know and 9\% believed that both pedestrians were extroverted. We believe that question $Q6$ caused a greater variety of perceptions from part of the participants due to the fact that we did not explain any concept when asking the questions, nor mentioned that the perceptions would be given from the geometric point of view, considering the position of the pedestrians in the space. Many of the participants, when questioned about extroversion, may have been influenced by the movements and appearances of the humanoids rather than the geometric features. In this sense, in question $Q7$, instead of which pedestrian was more extroverted, we asked which of the pedestrians appeared to be more sociable. When asked which pedestrian appeared to be more sociable, in question $Q7$, most of the participants (57\% of them) seemed to be more convinced that the pedestrian highlighted in yellow is the most sociable, in accordance with the model proposed by~\cite{favaretto2018detecting}.

\section{Final Considerations}
\label{sec:finalConsiderations}

This work evaluated people's perceptions with respect to geometric features, such as: density, speed, angular variation and distances among pedestrians. We also evaluated subjects perception regarding other subtle parameters as personalities and emotions traits in crowds. We proposed and implemented a survey that has been answered by 73 participants through a questionnaire that featured visualizations of scenes taken from videos of the \textit{Cultural Crowds}~\cite{favaretto2016using} dataset and propose questions regarding variation of visualization parameters.

Regarding the results of the people's perceptions about the geometric data, in the general analysis of the cameras, it was noticed that the way agents are displayed and the camera point of view interfere in the parameters perception. In particular, the greater the distance from the camera to the environment (oblique and top cameras), the better seem to be the perception of density, speed and angular variation. With respect to density, we can see that there was a more accurate perception in the first person view when the environment contained walls around the agents. Concerning speed parameter, subjects perceive better the speed variation of the avatars running through the oblique camera than in the top camera. 
In general analysis of the avatars type, there was a more accurate perception of density when visualized as humanoids in the first-person view, a better perception of angular variation through the humanoids in all the cameras, and more accurate perception of distances when avatars were displayed as cylinders in the top and oblique cameras.
We also performed an experiment to evaluate if people can perceive different personalities and emotions performed by pedestrians in crowds. 
It was interesting to see that, even without explaining to the participants the concepts of each personality or emotion and how they were calculated in our approach (considering the geometric characteristics), in all the cases, more than half of the participants perceived the personality and emotion that the agent was expressing in the video, in accordance with our approach. Of course, this last aspect is much more intangible and the missing explanations that we were interested about spatial manifestation and not trying to "figure out" if the person is social or open in a psychological point of view is certainly one aspect we want to deal in a future work.


%

\bibliographystyle{ACM-Reference-Format}
\bibliography{sample-base,refs}

\end{document}